\documentclass{ws-procs9x6}

\usepackage{floatflt}
\usepackage{graphics}
\usepackage{epsfig}

\newcommand{\be}{\begin{equation}}\newcommand{\ee}{\end{equation}}
\newcommand{\e}{\end{equation}}
\newcommand{\bea}{\begin{eqnarray}}
\newcommand{\eea}{\end{eqnarray}}
\newcommand{\beq}{\begin{eqnarray}}
\newcommand{\eeq}{\end{eqnarray}}
\def\({\left(}
\def\){\right)}
\def\[{\left[}
\def\]{\right]}
\def\vect#1{\skew{-1}{\mathaccent"017E}{#1}}
\def\vp{\vect p}

\def\tmp#1{}
\begin{document}

%\title{Casimir interaction between a perfect
%conductor and graphene described by the Dirac model}

%\title{On Casimir force between ideal conductor
%and suspended graphene modeled with relativistic fermions}

\title{Casimir energy of finite width mirrors:\\
renormalization, self-interaction limit and Lifshitz formula}

\author{I.~V.~Fialkovsky$^{\dag\ddag}$,
V. N.  Markov$^\P$ and  Yu. M. Pis'mak$^\ddag$
\footnote{The authors gladly acknowledge the support of
grants RNP 2.1.1/1575, RFBR $07$--$01$--$00692$. IVF is especially grateful
to support of University of Oklahoma and FAPESP.}
%\footnote{The authors gladly acknowledge the financial support of
%grants RNP 2.1.1/1575 and RFBR $07$--$01$--$00692$. IVF is especially grateful
%to continuous support of FAPESP.}
}
\address{
$^\dag$ ifialk@gmail.com, Instituto de F\'isica, Universidade de S\~ao Paulo, S\~ao Paulo, Brazil  \\
$^\ddag$Department of Theoretical Physics, Saint-Petersburg State University, Russia\\
$^\P$St. Petersburg Nuclear Physics Institute, Russia}

\begin{abstract}
We study the field theoretical model of a  scalar field in
presence of spacial inhomogeneities in form of one and two finite width mirrors
(material slabs). The interaction of the scalar field with the
defect is described with position-dependent mass term.
Within this model we derive the interaction of two finite width mirrors,
establish the correspondence of the model to the Lifshitz formula and
construct limiting procedure to obtain finite self-energy of a single mirror
without any normalization condition.
\end{abstract}
%\pacs{12.20.Ds, 73.22.-f}
\keywords{Casimir energy, QFT, finite width mirrors}
\bodymatter
%\vfill\eject
\section{Introduction}

The Casimir effect\cite{Casimir'48} was original considered for an extremely idealized configuration of electromagnetic field subject to ideal conducting boundary conditions on two parallel plates. Development of both theoretical and experimental techniques required consideration of more realistic systems which would describe real materials, both in their shape and properties. During last decades a number of such approaches were successfully developed, see \cite{Bordag-Mohideen-Mostepanenko-OBZOR'01} for a review.

One of the methods for investigation of (more) realistic situations was initially proposed by Symanzik \cite{Symanzik'81} and consists of modeling and generalizing the rigid boundary conditions by introducing additional singular-potential terms into the action of the model. This approach was widely explored for the case of delta-type potentials. The essence of this paper is to present the results of elaboration of similar field-theoretical approach in the case of step-potentials. The detailed calculations and discussions are presented~in \cite{FMPslabs,FMPComm08}.

\section{Single finite width mirror}

Let us consider a scalar field interacting with a
space defect with nonzero volume. Using the Symanzik's approach\cite{Symanzik'81}, we add
to the action an additional mass term being non-zero only inside the defect:
\be
\begin{matrix}
  S = S_0+S_{\it def} \\
         S_0=\frac12\int d^4x \phi(x) (-\partial^2_x+m^2) \phi(x) ,\quad
       S_{\it def}=\frac\lambda2\int d^4x  \theta(\ell, x_3)\phi^2(x)
       \label{action}
\end{matrix}
\ee
where $\partial_x^2=\partial^2/\partial
x_0^2+\ldots+\partial^2/\partial x_3^2$.
In the simplest case the defect could be considered as homogenous
and isotropic infinite plane layer of the thickness $\ell$ placed
in the $x_1x_2$ plane (so called `piecewise constant potential')
\be \theta(\ell, x_3)\equiv
    [\theta(x_3+\ell/2)-\theta(x_3-\ell/2)]/\ell.
    \label{theta}
\ee

To describe all physical properties of the systems it is
sufficient to calculate the generating functional for the Green's
functions
\be
    G[J]= N\int D\phi\, \exp\{-S[\phi]+J\phi\}
        \label{G(J)}
\e
where $J$ is an external source. The explicit calculation of $G[J]$
is performed with help of integral operators method developed by the authors, %for details
see \cite{FMP slabs,FMP Comm 08}.

\subsection{Casimir self-energy and its renormalization}
The Casimir energy density per unit area of the defect $S$ can be
presented with the relation
$ {\cal E}=-\frac{1}{TS}\ln G[0]
\label{E_TrLn}
$ , here $T$ is the (infinite) time interval  and $S$ --- the surface area of the defect.

For the case of a single finite--width mirror, as in (\ref{action}),
the Casimir energy is given by a sum of a UV finite, ${\cal E}_{\it fin}$,
and UV divergent, ${\cal E}_{\it div}$, parts
\be
    {\cal E}= {\cal E}_{\it fin} + {\cal E}_{\it div},
    \label{E_single}
\ee
 %${\cal E}_{\it div}$ is UV divergent in $d=4$
\be
{\cal E}_{\it fin}
    = \frac1{4 \pi^2}\int_0^\infty \( \ln\frac{e^{-\ell(E+\sqrt\rho)}}{4E\sqrt\rho\xi}
        -\frac{\lambda}{2E}\(1-\frac{\lambda}{4 \ell E^2}\)\)   p^2 dp , \quad %\label{E_fin}
        \ee
        $$
{\cal E}_{\it div}
    =\frac{\lambda\mu^{4-d}}{2 (2 \pi)^{d-1}}
    \int \frac{d^{d-1} p}{2E}\(1-\frac{\lambda}{4 \ell E^2}\),
    \label{E_div}
$$
where we used dimensional regularization, and put
$\rho=E^2+\kappa$, $\kappa={\lambda}/{l}$,
$ \xi^{-1}={e^{2\ell\sqrt{\rho}}(E+\sqrt{\rho})^2-(E-\sqrt{\rho})^2}$.
With appropriate redefinitions of
the parameters this expression coincides with known results in the literature \cite{um_slab}.

For renormalization of the model  at the one-loop level considered
here we must add to the action  the following
field-independent counter-term $\delta S$
\be
    \delta S = f \lambda   + g \lambda^2 \ell^{-1},
    \label{c-terms}
\ee
with bare parameters $f$ and $g$. %(of mass dimensions two and zero correspondingly).
Within such `minimal addition' renormalization scheme we
obtain for the renormalized Casimir energy %the following result
\be
    {\cal E}_r= {\cal E}_{\it fin} + \lambda f_r + g_r {\lambda^2 }{\ell}^{-1}
    \label{E_ren}
\e
where finite  parameters $f_r$, $g_r$ must be determined with
appropriate experiments, or fixed with normalization conditions.
The number of required conditions is dictated by the
(in)dependence of the coupling constant $\lambda$ on the slab
thickness $\ell$.

\subsection{Dirichlet limit and normalization condition}
Under the change of variables  $\lambda\to\ell \kappa$  two counter-terms
in (\ref{E_ren}) can be effectively combined into a single one
$\tilde g_r$ of mass dimension one \be
    {\cal E}_r= {\cal E}_{\it fin} + \kappa \ell \tilde g_r
    \label{E_ren1}
\e
One notes that putting $ \kappa=-m^2 $
and taking the $m\to\infty$ limit effectively converts
the system under consideration into a massless scalar field
confined between two plates at $x_3=\pm\ell/2$ 
%$x_3\in(-\ell/2,\ell/2)$ 
subject to Dirichlet boundary conditions
at the boundaries. We can use this correspondance for fixing $\tilde g_r$.

The finite part of the Casimir
energy (\ref{E_single}) in the  limit $m\to \infty$ yields
\be
   {\cal E}_{\it fin}=-\frac{m^4\ell}{128 \pi^2}
        +\(\frac\pi{6}-\frac{4}{9}\)\frac{m^3}{4 \pi^2}
        -\frac{\pi^2}{1440 \ell^3}+O(m^{-1})
   \label{E fin Dir}
\ee
Now we require that the renormalized energy (\ref{E_ren1}) in
this limit lead to the same Casimir pressure as in the case of
massless scalar field subject to Dirichlet boundary conditions
$
    {\cal P}_{\it Dir}=-\frac{\pi^2}{480 \ell^4}.
$
Then this condition fixes the renormalization parameter $\tilde g_r$ of
(\ref{E_ren1})
\be
    \tilde g_r=-\frac{m^2}{128 \pi^2}
    \label{t g_r}
\ee
Thus, using this Dirichlet limit procedure   we are
able to collate a particular limit of our results with a well known (unambiguous) physical situation.

\section{Interaction of two finite width mirrors}
Let us consider two plane slabs of thickness
$\ell_{1,2}$ interacting over the distance $r$.  The action can be written as
\be
\begin{matrix}
    S =S_0+S_{\it def}  \label{action 2L} , \quad
     S_0=\frac12\int d^4x \phi(x) (-\partial^2_x+m^2) \phi(x)
 \\
     S_{\it def}=
        \int d^3x \(\kappa_2\int_{-a_2-l_2}^{-a_2} dx_3 \phi^2(x)
            +\kappa_1\int_{a_1}^{a_1+l_1} dx_3\phi^2(x)\)
 \\
\end{matrix}
\ee
here  $a_1+a_2\equiv r$.

Proceeding along the lines of \cite{FMP slabs,FMP Comm 08} we get the final expression for the energy
\be {\cal E}_{2L}
    ={\cal E}_{1}+{\cal E}_{2}
    +\int\frac{d^3\vp}{2(2\pi)^3}
        \log
        \[1-e^{-2 E r}\prod_{i=1,2}\kappa_i\xi_i
                (1-e^{2 \ell_i\sqrt{\rho_i}})
        \].
    \label{E 2L}
\ee
Here ${\cal E}_{1,2}$ give the self-energy (\ref{E_single}) of solitary layers
$1$, $2$ correspondingly.
All the notation here follows ones of Sect. 2 with subscript index corresponding to
the layer number.

The third term in (\ref{E 2L}) represents
the interaction of two layers and vanishes in the limit
$r\to\infty$. We note that the interaction term is UV finite, and
the removal of regularization made in (\ref{E 2L}) is indeed
justified. This is in perfect accordance with general
considerations \cite{Emig} of the finiteness of Casimir interaction
between disjoint bodies.

Basing on the general expression (\ref{E 2L})
we can calculate the vacuum energy in different limits
such as self-pressure of the slab in presence of delta-spike and interaction between two of them.
The two delta-spikes limit known in previous literature  \cite{} is also reproduced.

\subsection{Connection to the Lifshitz formula}
Now we consider in (\ref{E 2L}) the limit of slabs of infinite width separated by
finite distance $r$. For the force in such a limit we obtain
\be
  {\cal F}_{\it Lif} \equiv
    -\frac{\partial {\cal E}_{\it Lif}}{\partial r}
    =-\int\frac{d^3\vp}{(2\pi)^3}
        \frac{E}
        {e^{2 E r} \frac{(E+\sqrt{\rho_1}) (E+\sqrt{\rho_2})} {(E-\sqrt{\rho_1}) (E-\sqrt{\rho_2})}-1}
        \label{F-lif}
\ee
It is straightforward to see that the correspondence with the Lifshitz formula \cite{Lifshitz'56}
achieved if we introduce particular dispersion
into the interaction of quantum fields with the material defect,
\be
    \kappa_{1,2}^{TE}(p)=(\epsilon_{1,2}-1) p_0^2,  \, \, \,
    \kappa_{1,2}^{TM}(p)
    =\(\frac{1}{\epsilon_{1,2}^2}-1\) \vec{p}^2+\(\frac{1}{\epsilon_{1,2}}-1\) p_0^2.
\ee
The parameter $\epsilon$ which enters these dispersion relations
is to be identified with dielectric permittivity.
Summing TE and TM  contributions we immediately recover the Lifshitz formula \cite{Lifshitz'56}
\be
  {\cal F}_{\it Lif} =
    -\frac{1}{4 \pi^2} \int_0^{\infty} dp_0 \int_0^{\infty} d p^2 E (d_{TE}^{-1}+d_{TM}^{-1})
\ee
\be
   d_{(TE,TM)}   =  {e^{2 E r} \frac{(E+\sqrt{\rho^{(TE,TM)}_1}) (E+\sqrt{\rho^{(TE,TM)}_2})} {(E-\sqrt{\rho^{(TE,TM)}_1}) (E-\sqrt{\rho^{(TE,TM)}_2})}-1}
\ee
with $\rho^{(TE,TM)}_{1,2}=E^2+\kappa^{(TE,TM)}_{1,2}(p)$.

\subsection{Self-interaction limit}
For $\kappa_2=\kappa_1$ the limit $\ell\to\infty$
taken in the two-slab action (\ref{action 2L}) reproduces the single-slab action
(\ref{action}) subject to the substitution $\ell\to r$, $m^2\to m^2+\kappa_1$,
$\lambda\to-\kappa_1 r$. One would expect that the Casimir energy of two slabs in this limit also
reproduces the result for a single finite-width mirror.

However, one finds that $ {\cal F}_{\it Lif}$ (\ref{F-lif}),
being UV finite in this limit, differs from UV divergent $r$-derivative of (\ref{E_single})
${\partial \cal E}/{\partial r}$ by
\be
  \Delta = %\frac{\partial \cal E}{\partial r}-\frac{\partial \cal E_{\it Lif}}{\partial r}
    -\frac{\mu^{4-d}}{2(2\pi)^{d-1}}\int d^{d-1}\vp
            \(E-\sqrt{E^2+\kappa}\)
\ee
Yet, this discrepancy %reflects the very nature of presented calculations but
does not signal any inconsistency.
In Sec. 2 we presented a derivation of the Casimir
\textit{self-}energy of a single slab. It is known that in general
the self-interactions of Casimir type do possess divergencies
depending on the geometrical properties of the system. This fact
is reflected in the presence of the counter terms (\ref{c-terms}).

On the other hand, it is well known \cite{Emig},
that Casimir interaction of two \textit{disjoint} bodies is always free of
divergencies depending on the distance between them. This
very system was considered in this section, and the force (\ref{F-lif})
between two distinct slabs was found to be finite
and %perfectly
unambiguous. Consequently, the limit of infinite width of the slabs which %formally
recover the self-pressure of a single body, does not bring any divergencies or ambiguities.

Requiring that (\ref{E_ren1}) coincides with (\ref{F-lif}) we can fix the counter terms $\tilde g_r$
and collate the two approaches.
One finds that this condition fixes the same value for
$\tilde g_r$  as one elaborated in Sec. 4 and given by (\ref{t g_r}).

Thus, we establish a new divergence free approach to calculate the self-pressure of a single finite width slab
which lead to the same result as one obtained %in a standard way upon
imposing a physically
motivated normalization condition.

\bibliographystyle{ws-procs9x6}
%\bibliography{ws-pro-sample}

\end{document}